\PassOptionsToPackage{prologue,table,xcdraw}{xcolor}
\documentclass[sigconf,screen]{acmart}

\AtBeginDocument{%
  }

\usepackage{amsmath,amsfonts}
\usepackage[ruled,vlined,linesnumbered]{algorithm2e}
\usepackage{graphicx}
\usepackage{textcomp}
\usepackage{cleveref}
\usepackage{url}
\usepackage{xspace}
\usepackage{multirow}
\usepackage{enumitem}
\usepackage{pifont}
\usepackage[table,xcdraw]{xcolor}

\theoremstyle{definition}
\newtheorem{definition}{Definition}[section]

\usepackage{framed}
\newcommand{\mybox}[1]{
    \setlength{\FrameRule}{0.4pt} 
    \setlength{\FrameSep}{5pt}    
    \begin{framed}
        \setlength{\parskip}{0pt}  
        \setlength{\parindent}{0pt}
        #1
    \end{framed}
}

\setcopyright{rightsretained}
\copyrightyear{2026}
\acmYear{2026}
\acmDOI{10.1145/3744916.3773233}
\acmConference[ICSE '26]{2026 IEEE/ACM 48th International Conference on Software Engineering}{April 12--18, 2026}{Rio de Janeiro, Brazil}
\acmBooktitle{2026 IEEE/ACM 48th International Conference on Software Engineering (ICSE '26), April 12--18, 2026, Rio de Janeiro, Brazil}
\acmPrice{}
\acmISBN{979-8-4007-2025-3/26/04}
\acmSubmissionID{icse2026-p4111}

\begin{document}

\title{{Minimizing Breaking Changes and Redundancy in Mitigating Technical Lag for Java Projects}}

\author{Rui Lu}
\authornote{Rui Lu and Lyuye Zhang contribute equally.}
\affiliation{%
  \institution{Shanghai Key Laboratory of Trustworthy Computing, East China Normal University}
  \city{Shanghai}
  \country{China}
}
\orcid{0009-0006-3311-0590}
\email{ruilu@stu.ecnu.edu.cn}

\author{Lyuye Zhang}
\authornotemark[1]
\affiliation{%
  \institution{Nanyang Technological University}
  \city{Singapore}
  \country{Singapore}
}
\orcid{0000-0003-3087-9645}
\email{zh0004ye@e.ntu.edu.sg}

\author{Kaixuan Li}
\authornotemark[2]
\affiliation{%
\authornote{Kaixuan Li and Min Zhang are corresponding authors.}
  \institution{Shanghai Key Laboratory of Trustworthy Computing, East China Normal University}
  \city{Shanghai}
  \country{China}
}
\orcid{0000-0002-3517-353X}
\email{kaixuanli@stu.ecnu.edu.cn}

\author{Min Zhang}
\authornotemark[2]
\affiliation{%
  \institution{Shanghai Key Laboratory of Trustworthy Computing, East China Normal University}
  \city{Shanghai}
  \country{China}
}
\orcid{0000-0002-3152-4347}
\email{mzhang@sei.ecnu.edu.cn}

\author{Yixiang Chen}
\affiliation{%
  \institution{Shanghai Key Laboratory of Trustworthy Computing, East China Normal University}
  \city{Shanghai}
  \country{China}
}
\orcid{0000-0003-1235-5530}
\email{yxchen@sei.ecnu.edu.cn}



\newcommand{\tool}{\textsc{DepUpdater}\xspace}

\begin{abstract}
Re-using open-source software (OSS) can avoid reinventing the wheel, but failing to keep it up-to-date can lead to missing new features and persistent bugs or vulnerabilities that have already been resolved. The use of outdated OSS libraries introduces technical lag, necessitating timely upgrades. However, maintaining up-to-date libraries is challenging, as it may introduce {incompatibility issues} that break the project or {redundant dependencies that unnecessarily increase the size of the project}. These issues discourage developers from upgrading libraries, highlighting the need for a fully automated solution that balances version upgrades, reduces technical lag, ensures compatibility, and avoids redundant dependencies.

To this end, we propose \tool, which ensures that upgrades minimize technical lag as much as possible while avoiding incompatibility issues and redundant dependencies. The comparison with existing dependency management tools demonstrates that \tool more effectively reduces technical lag while ensuring compatibility and pruning redundant dependencies.
Additionally, an ablation study highlights the potential benefits of considering {pruning} requirements during upgrades to mitigate {incompatibility issues}. Finally, leveraging \tool, we investigate the impact of transitive dependency upgrades on client compatibility, providing insights for future research.
\end{abstract}

\maketitle

\section{Introduction}
In software development, open-source software (OSS) is widely used to save developers time and effort. The software projects that use OSS libraries are considered {client projects}, with the libraries serving as {dependencies}. 
Along with the development and evolution of client projects, the dependencies may not be up-to-date. 
This results in client projects missing out on the improved features provided by the new versions of their dependencies. More critically, some libraries address security vulnerabilities during upgrades. If the client project does not upgrade dependencies in time, it remains vulnerable to these security risks~\cite{pashchenko2020qualitative, wu2023understanding,cox2015measuring,prana2021out, pashchenko2018vulnerable,zhang2025fixing}.

Given the widespread presence of outdated dependencies and the risks they pose, many research studies have revealed this problem.
González-Barahona et al.~\cite{gonzalez2017technical} introduced the concept of technical lag to reflect the degree to which a software deployment relies on {outdated libraries. Existing studies on technical lag}
~\cite{gonzalez2020characterizing, zerouali2019formal, wang2020empirical, decan2018evolution, stringer2020technical, gonzalez2017technical} highlight that it is a practical metric for measuring the outdateness of dependencies.
Typically, upgrading outdated dependencies to reduce technical lag is an effective way to maintain project quality and mitigate potential risks,
and the latest versions of dependencies generally have fewer known vulnerabilities~\cite{huang2022characterizing}. 
However, upgrading is not a trivial task. The upgrade process may introduce breaking changes that cause software failures in the client projects, known as \textbf{incompatibility issues}~\cite{dig2006apis, dietrich2014broken, xavier2017historical, jayasuriya2023understanding, jayasuriya2024understanding}. 
Therefore, software developers are reluctant to upgrade their dependencies~\cite{zhang2023compatible}. 

In addition to {incompatibility issues}, upgrading dependencies without careful consideration can sometimes introduce unnecessary dependencies not required for the project’s build or runtime. These dependencies increase the overall size of the project with escalated maintainability burden and the risk of being attacked by supply chain attacks~\cite{song2024efficiently, quach2018debloating, vazquez2019slimming, rastogi2017cimplifier, heo2018effective, qian2019razor,sharif2018trimmer,soto2021comprehensive,ladisa2023sok,zhao2023fse}. 
We define the newly introduced dependencies as \textbf{redundant dependencies}, which are not necessary for the client because they were not in use before upgrading. The technique to avoid redundant dependencies is \textbf{pruning}.

Incompatibility issues and redundant dependencies are significant obstacles to effectively upgrading dependencies for reducing technical lag. However, existing tools~\cite{dependabot, snyk, jaime2024balancing} do not consider both factors simultaneously during the upgrade process.
To bridge this gap, the following objectives have to be achieved: 
(1) Fully mitigating technical lag by upgrading all dependencies of client projects, including both direct and transitive dependencies. (2) Preventing incompatibility issues that could break the client. (3) Avoiding the introduction of {redundant} dependencies.

Given the complexity of dependency management, three major challenges must be addressed to achieve the objectives: \textbf{C1}: Upgrading a single library may alter the dependency relationship and affect the compatibility of other dependencies.
\textbf{C2}: Compatibility assessment should be based on the client's perspective of API usage.
\textbf{C3}: The determination of whether dependencies are {redundant} should be based on the structure of the current dependency graph, and the graph often has a dynamic structure.

To address these challenges, we propose \tool, which minimizes technical lag while preventing syntactic incompatibility issues and avoiding the introduction of redundant dependencies.
\tool follows a systematic order to upgrade all dependencies in the dependency graph and updates the impact of each upgrade to address \textbf{C1}. Then, \tool performs a points-to analysis (PTA) to track APIs and Class references by the client, both directly and indirectly, allowing the evaluation of the actual impact on the client project to address \textbf{C2}. Finally, \tool tracks the transitive dependencies introduced by each version, combined with the current structure of the dependency graph, to filter out versions that would introduce {redundant} dependencies if upgrades to these versions, addressing \textbf{C3}.
We selected Maven as an example due to its rapid version iterations, making it more prone to technical lag, as supported by a recent report showing Maven projects have the largest average released versions (28 versions) compared with NPM~\cite{npm} (10) and Pypi~\cite{pypi} (10)~\cite{sonatypeStudy}. Our approach, however, is applicable to other version-based platforms like NPM.

To comprehensively evaluate the effectiveness and insights of our approach, we conducted three experiments:
\ding{172} RQ1: We compared \tool against four state-of-the-art baselines, demonstrating that \tool significantly outperforms existing tools in mitigating technical lag.
\ding{173} RQ2: We systematically removed two key components of \tool, avoiding incompatibility issues and redundant dependencies, individually and jointly. Results show that each component contributes substantially to the overall performance.
\ding{174} RQ3: Using a larger dataset of 1,529 modules, we investigated the real-world impact of transitive dependency upgrades. The findings reveal that while such upgrades may introduce breakages, dependencies located beyond six layers deep in the dependency tree are less likely to cause client failures.
The dataset used in RQ1 and RQ2 includes 356 modules with full compilation and testing coverage. RQ3 leverages a larger dataset with 1,529 modules. Both datasets were randomly sampled from top-starred Maven projects on GitHub to reflect the practices and trends of widely used Java projects.
Our contributions are:
\begin{itemize}[leftmargin=5pt]
    \item We proposed \tool as a dependency management tool for Maven projects to minimize technical lag while ensuring  {syntactic compatibility} and avoiding {redundant} dependencies.
    \item We identified the potential contribution of {pruning} in facilitating compatible upgrades {for dependencies}.
    \item We obtained the distribution of client-impacting APIs and broken clients across dependency depths when upgrading transitive dependencies using trivial strategies, providing a reference for researchers in this field.
\end{itemize}
Practically, \tool can be integrated into both the maintenance phase and the development phase to automatically upgrade outdated dependencies without breaking the projects. We have open-sourced \tool and the experiment data
\footnote{\url{https://github.com/ruisearch/DepUpdater}}.

\section{Background}
\subsection{Technical Lag}
Modern software relies on third-party libraries managed by package managers like Maven~\cite{maven}, NPM~\cite{npm}, and Cargo~\cite{cargo}. 
Since libraries evolve independently, the libraries used by projects often remain locked to older versions, missing new features and bug fixes, leading to technical lag~\cite{gonzalez2017technical,stringer2020technical}. Zerouali et al.~\cite{zerouali2019formal} proposed two methods to measure technical lag: one based on time (release date differences) and another on version numbers (using Semantic Versioning ~\cite{semver}). 
\subsection{Dependency Management by Maven}\label{sec:2.2}
Maven~\cite{maven} is a widely used package manager for Java projects, where dependencies are managed through the Project Object Model (POM) file. As its representative, following previous work~\cite{jayasuriya2023understanding,dann2023upcy,zhang2023compatible,jaime2024balancing}, we focus on Java projects managed by Maven. For these projects, developers specify dependency versions, and Maven resolves and downloads them from the MCR~\cite{MavenCentral}. Dependencies explicitly declared in the POM file are direct dependencies, while others pulled in transitively are transitive dependencies. Each dependency is uniquely identified by GAV coordinates~\cite{gav}; notably, the version is recommended to follow the rule of Semantic Versioning (SemVer)~\cite{semver}. Maven constructs a dependency graph, a directed acyclic graph (DAG) where nodes represent libraries and edges denote dependency relationships. Conflicts arise when multiple versions of a library exist, and Maven resolves them through dependency mediation, selecting the version from the shortest path to the client project~\cite{conflict}. Additionally, Maven transforms the graph into a dependency tree by shadowing dependency relationships. In Maven projects, it is common to have multiple POM files, each corresponding to a {module}, which can be built separately~\cite{module}.

\subsection{Compatibility}
Upgrading a dependency may introduce API changes that cause {incompatibility issues}, categorized as syntactic (compilation or linking failures due to API changes) and semantic (behavioral changes despite successful compilation)~\cite{jayasuriya2023understanding,jayasuriya2024understanding}. 
Following previous works~\cite{zhang2023compatible, jaime2024balancing, dann2023upcy,zhang2022has}, 
we use static analysis tools to detect {incompatibility issues}. Static analysis tools, which are widely used in incompatibility detection, such as Revapi~\cite{revapi}, target syntactic incompatibility issues and are not sufficient for detecting semantic issues, which are usually caught by regression testing~\cite{greca2023state,zhang2024mitigation}.
According to a recent study by Alex et al.~\cite{gyori2018evaluating}, syntactic {incompatibility issues} are more common than semantic {incompatibility issues} and have more impact on software. A study by Danny et al.~\cite{dig2006apis} also found that 80\% of the incompatibility issues are syntactic incompatibility issues.
Thus, our work mainly aims to avoid syntactic breaking issues and reduce the risks of semantic breaking as much as possible during the process of upgrading dependencies. 
{\subsection{Motivating Example}\label{sec:2.4}
Taking \textit{spring-cloud-alibaba}~\cite{alibaba}
as an example, it is a multi-module project with \textit{spring-cloud-starter-stream-rocketmq}~\cite{rocketmqmodule} as one of the modules. \tool identified that this module contains 120 outdated dependencies. However, manually upgrading all of them would incur substantial time and effort. Furthermore, further testing the feasibility of upgrading 120 dependencies sequentially is not realistic. Therefore, a tool is needed to automatically detect outdated dependencies in a project and recommend appropriate version upgrades for all dependencies with a holistic approach.
}
\section{Methodology}
\subsection{Overview}\label{sec:3.1}

\begin{figure*}[htbp]
    \centering
    \begin{minipage}{0.65\textwidth}
    \centering
    \includegraphics[width=\linewidth]{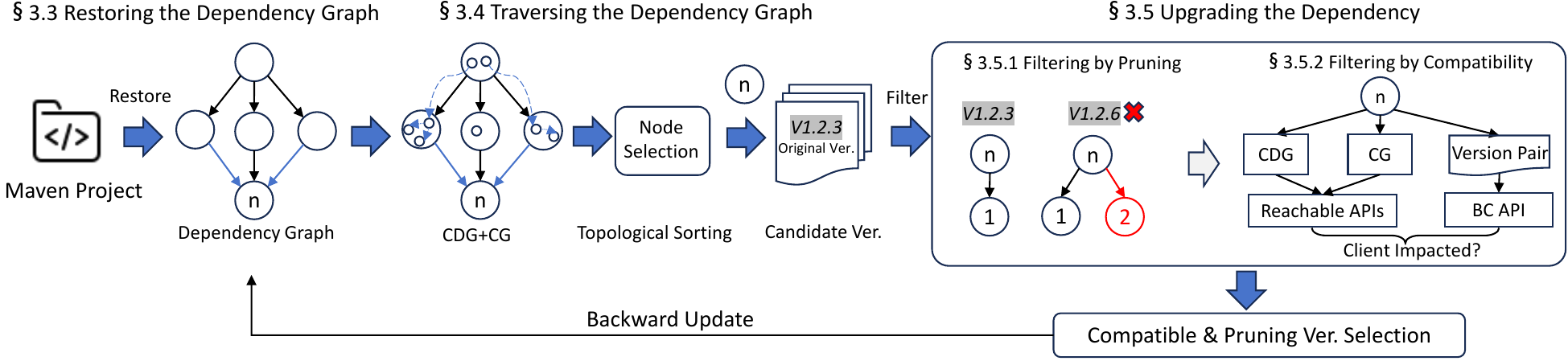}
    \caption{Overview of \tool.}
    \label{fig:tool}
    \end{minipage}
    \hfill
    \begin{minipage}{0.3\textwidth}
    \includegraphics[width=\linewidth]{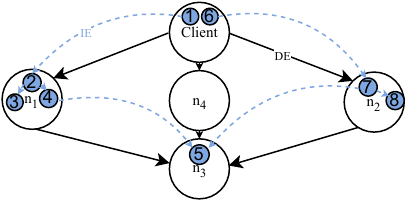}
    \caption{A Dependency Graph.}
    \label{fig:DG}
    \end{minipage}
\end{figure*}

To reduce technical lag effectively, we need to understand all the dependency relationships, which are typically represented in the dependency graph. Although the dependency graph is a classic concept in dependency management, unlike the existing work~\cite{jaime2024balancing, dann2023upcy, ochoa2022breaking,ochoa2021breaking, zhang2023compatible}, we define a different model of dependency graph used by \tool for the following reasons:
\begin{itemize}[leftmargin=5pt]
    \item \textbf{API Usage Analysis}: To accurately determine whether a change in an API will cause {incompatibility issues}, it's crucial to know whether that API is used. This requires detailed information about API invocations, which traditional dependency graphs do not include. By incorporating PTA, we can represent method and class invocation relationships in the dependency graph, enabling more precise compatibility analysis by identifying whether an API is invoked by the client code.
    \item \textbf{{Pruning Requirement}}: {Pruning} focuses on avoiding the {redundant} size of the dependency graph itself. To achieve this, we need to capture additional information in the graph. Specifically, our graph must include upstream dependencies for each potential upgrade candidate, as well as the subgraphs created by these dependencies. By identifying these subgraphs, the graph can show how upgrading a node introduces new dependencies that may connect with other parts of the graph. This information allows us to manage the complexity without isolating the subgraphs for independent calculation. 
\end{itemize}

As shown in~\Cref{fig:tool}, \tool is designed to reduce the technical lag of modules by upgrading all dependencies to their newest versions, which do not introduce {incompatibility issues} or redundant dependencies. To achieve this, the tool first constructs the above-mentioned graph based on the dependency graph derived from Maven. It then traverses this dependency graph, and for each node encountered, filters out versions that introduce incompatibility issues or redundant dependencies. The process is illustrated in \cref{sec:3.5.1,sec:3.5.2}.
{To detect incompatibility issues, we use static analysis to identify client-impacting APIs and compare them with breaking changes reported by Revapi. To detect redundant dependencies, we check whether an upgrade would introduce new dependencies absent from the original graph.}
From the remaining versions, the latest one is selected to minimize the technical lag of the module. After processing each node, the tool also updates the dependency graph in real-time, i.e., the structure of the dependency graph is promptly adjusted as dependencies are upgraded.

\subsubsection{Construction of Pre-computed Database}\label{sec:3.1.1}
{\tool uses a pre-computed and continuously updating local MongoDB database~\cite{MongoDB} to support the above-mentioned process. The database contains metadata about all versions and dependency relationships on~\textbf{Maven central repository (MCR)}~\cite{MavenCentral} to retrieve the candidate versions when handling a dependency, as well as updating the dependency graph after upgrades.
Specifically, we crawled the metadata from the MCR feeds, resulting in 15,399,676 versions, and then downloaded and parsed their POM files to obtain the dependency relationships with Maven commands. 12,746,588 relationships were derived.
} 
Building the local database took around 80 man-hours with a cut-off date of July 2024.

\subsection{Dependency Graph and Technical Lag}
\subsubsection{Dependency Graph}
The dependency graph that encapsulates all necessary information about the dependency relationships is a core concept in dependency management. 
For further reference and notation, we define the dependency graph model used in \tool as follows.

\begin{definition}[Dependency Graph]
\label{def:DG}
A \textit{dependency graph} $G$ = 
<$N$, $D$, $A$, $I$>
is a DAG where, 
\begin{itemize}[leftmargin=5pt]
\item $N$ as a set of nodes in the graph can be divided into two types: 
\begin{enumerate}
    \item \textbf{Client Project Node}: There is exactly one client project node. This node serves as the root of the graph. This node is denoted as $Client$.
    \item \textbf{Dependency Nodes}:  These represent the dependencies of the client project, including both direct dependencies (those directly referenced by the client project) and transitive dependencies (those referenced through other dependencies).
\end{enumerate}
Nodes in the graph have the following three properties:
\begin{enumerate}
    \item $Reachable: N\rightarrow A$ is a function that assigns the reachable APIs of a node in the graph. For $n\in N$, $ reachable (n)$ is a set of API in $n$ that are invoked directly or indirectly by the APIs in $Client$, denoted as the reachable APIs of $n$.
    \item $Version: N\rightarrow \mathcal{P}(V)$ is a function that assigns the candidate versions to a node. The candidate versions are those that the node can be upgraded to, consisting of the original version of the node and all versions that are newer than the original version. The $V$ is a set consisting of the candidate versions, and $\mathcal{P}(V)$ is the power set of $V$.
    \item $Upgrade: N\rightarrow V$ is a function that {assigns the upgraded version to a node}, where $V$ is the set of candidate versions. 
\end{enumerate}

\item $D\subseteq N\times N$ is a set of directed edges in the graph, representing dependency relationships, denoted as \textbf{Dependency Edge (DE)}. $ (u,v)\in D$ means node $u$ is a dependent and node $v$ is its direct successor in the graph.
\item $A$ is a set of {APIs}. There are two types of APIs: method and class.
\item $I\subseteq A\times A$ is a set of directed edges representing the API invocation relationship, denoted as \textbf{Invocation Edge (IE)}. $(a,b)\in I$ means API $a$ invokes API $b$ directly.
\end{itemize}
\end{definition}
Based on the definition above, consider a dependency graph shown in \Cref{fig:DG}. 
In this graph, nodes $n_1$ and $n_2$ are direct dependencies of the client project node (denoted as $Client$), and node $n_3$ is a transitive dependency, meaning it is required directly by $n_1$ and $n_2$, but not directly required by $Client$. 
Solid arrows represent the Dependency Edges $DE$. 
The numbers $1$ to $8$ in the graph represent APIs within the nodes, with dashed arrows representing the Invocation Edges $IE$.

\subsubsection{Technical Lag}
{The computation of technical lag is based on the dependency graph, because the project's technical lag is the sum of the lags of all dependencies in the graph~\cite{zerouali2019formal}. Although transitive dependencies are less likely to impact the client, the libraries included transitively are also more likely to be vulnerable~\cite{lauinger2018thou}. Therefore, we do not differentiate the way of computing the technical lag of direct dependencies and transitive dependencies. To clearly demonstrate the difference, we separately calculate the technical lag reduced on direct and transitive dependencies, respectively, in the evaluation ~\cref{sec:4.5,sec:4.6}. Following works on technical lag~\cite{gonzalez2017technical,zerouali2019formal}, there are two ways to compute the technical lag among a project: version lag and time lag~\cite{zerouali2019formal, cox2015measuring, stringer2020technical,decan2018evolution}.}
Based on the dependency graph defined in \Cref{def:DG}, we define the technical lag used in \tool and its evaluation as follows:

1) \textbf{Time Lag}:
The concept of {time lag} is defined in a manner consistent with previous work~\cite{gonzalez2020characterizing}. It quantifies the delay in dependency upgrades based on release dates. Specifically, for each dependency in a project, its time lag is measured as the number of days between the release date of the currently used version and the release date of the latest available version. 
The total time lag of a project is then computed as the sum of the \texttt{time lag} of all its dependencies, including direct ones and transitive ones.

\begin{definition}[Time Lag]
\label{def:TL}
{Given a dependency graph $G$ = <$N$, $D$, $A$, $I$> and its client project $Client$,}
for each dependency $d\in N\backslash Client$,
let $v(d)\in version(d)$ be its current version and $v^*(d)\in version(d)$ be the latest available version, {where the ``latest'' refers to the version with the most recent release time}.
The time lag of $d$ is $lag_t(d) = \operatorname{days}(\operatorname{release}(v^*(d))
- \operatorname{release}(v(d)))$, where $release(v)$ denotes the release date of version $v$, and $days(\cdot)$ is a function that converts the time difference into days.
The \textbf{time lag} of the 
client project (denoted as $Client$ in $G$) is then given by:
\begin{align}
TL(Client) &= \sum_{d \in N\backslash Client} lag_t(d) \nonumber \\
      &= \sum_{d \in N\backslash Client} \operatorname{days}(\operatorname{release}(v^*(d)) - \operatorname{release}(v(d))) \nonumber
\end{align}
\end{definition}

2)  \textbf{Version Lag}:
In this paper, technical lag is measured as the distance between the versions of deployed dependencies and the latest available versions. Versions are sorted based on SemVer, meaning they are sorted by major, minor, and then patch version, aligning with how \tool sorts versions. 
\begin{definition}[Version Lag]
\label{def:VL}
{Given a dependency graph $G$ = <$N$, $D$, $A$, $I$> and its client project $Client$,} 
for each dependency $d\in N\backslash Client$, let $v(d)\in version(d)$ be its current version and $v^*(d) \in version(d)$ be the latest available version, where the ``latest'' refers to the version that has the largest version number based on SemVer~\cite{semver}.
Then, the version lag of a dependency $d$ is denoted as 
$lag(d) = \lvert \{ v' \mid v' > v(d), v' \text{ is a stable version of } d \} \rvert$. 
That is, $lag(d)$ counts the number of stable versions of $d$ that have a larger version number than $v(d)$ based on SemVer.
The \textbf{version lag} of the client project $Client$, 
denoted as $VL(Client)$, is the total lag accumulated across all dependencies in $N\backslash Client$:
\begin{align}
\label{VL}
VL(Client) &= \sum_{d \in N\backslash Client} lag(d) \nonumber\\
      &= \sum_{d \in N\backslash Client} \left| \{ v' \mid v' > v(d), v' \text{ is a stable version of } d \} \right| \nonumber
\end{align}

\end{definition}
{Following prior works~\cite{zerouali2019formal,gonzalez2017technical}, we measure version lag not only as the total number of versions between the deployed and latest versions, but also at the major, minor, and patch levels, since changes at different levels imply different upgrade risks. Major-level version lag refers to upgrades that increase the major number, minor-level version lag to those that increase the minor number while keeping the major the same, and patch-level version lag to those that increase only the patch number. Upgrades that do not increase any of the three numeric digits are classified as pre-release level version lag.
For example, given version 1.0.1, upgrading version 1.0.1 to version 2.0.0 reduces major-level version lag because it increases the major number from 1 to 2.
}


\subsection{Restoring the Dependency Graph}
The dependency tree displays the versions of dependencies resolved by the Maven project. \tool parses the tree to retrieve version information. As mentioned in~\Cref{sec:2.2}, this tree does not reflect the versions of all dependencies in the Maven project, because Maven prunes the dependency graph into a tree by shadowing edges in the case of duplicates or conflicts.
However, to detect client-impacting APIs, it is necessary to restore these hidden relationships. \tool achieves this by using the -Dverbose option~\cite{Dverbose} to include omitted dependency relationships, reconstructing the complete dependency graph. {To focus on relevant dependencies, \tool excludes those with test or provided scope, }{and only considers the dependencies with compile or runtime scope. Because the dependencies with test or provided scope do not provide functionality in the deployment and runtime process of the client, we exclude them following~\cite{zhang2023mitigating,zhang2023compatible}.}

In multi-module projects with hierarchical POM files, it is common for one local module to depend on another, and the artifacts of these modules may not be available in the MCR. To manage these local dependencies, \tool scans the project’s repository structure to identify local modules that are not published to the central repository. When processing a module, if a dependency on another local module is detected, \tool constructs the local module locally instead of attempting to download it from the MCR. By considering the entire project structure rather than processing modules one by one, \tool ensures that all local dependencies are built in advance, preventing potential issues with unresolved dependencies. Handling modules individually would not effectively resolve cases involving local modules, as these local modules need to be built in advance for the convenience of proper analysis of their code.

\subsection{Traversing the Dependency Graph}
Through the restored graph, all the dependency relationships are clear. \tool uses this information to achieve the primary objective: Updating dependencies to reduce the technical lag without breaking the client project or introducing redundant dependencies.
To minimize technical lag as much as possible, \tool upgrades each dependency in the graph to the optimal version after obtaining the dependency graph. {It is not suitable to upgrade all nodes at once, 
as upgrading a single node can alter dependency relationships and affect the reachable APIs of other nodes, resulting in a dynamic dependency graph during the upgrading process. Without a proper traversing order, the retrospective effect of past upgrades may contradict the current upgrade, which requires constant backtracking to traverse the most optimal configuration, leading to excessive cost. Given this dynamic structure, it is necessary to upgrade nodes in a specific order and update the graph accordingly. This is the reason why those tools that upgrade one dependency at a time are not applicable for minimizing technical lag.}

We design a novel algorithm in \tool to traverse the dependency graph in order, which is similar to the topological sorting algorithm~\cite{Topo}. This design is based on the following rationales: 
\begin{itemize}[leftmargin=5pt]
\item \textbf{Structure of Dependency Graph}: The dependency graph is a DAG, so the topological sorting algorithm is an applicable way to traverse all nodes in order;

\item \textbf{Prerequisite for Updating a Dependency}: 
{\tool uses PTA with Soot~\cite{sootspark} to analyze bytecode, tracking method calls, and class references, {which refer to the type of Java classes that are referenced~\cite{javaspec},} across direct and transitive dependencies. This enables precise detection of API usage and breaking changes. In contrast, prior works like CORAL~\cite{zhang2023compatible} and Steady~\cite{steady} relied on call graph analysis, which misses cases where removed classes affect return values without direct method calls. }To ensure accurate impact analysis, \tool updates all dependents before evaluating a dependency, allowing PTA to track reference flows correctly. It employs topological sorting to maintain a consistent {upgrade order}, ensuring that each node is processed only after its predecessors have been processed. 
{The topological sorting algorithm can ensure that the calculation of each node in the dependency graph follows a forward-only traversal, thanks to the nature of the Maven acyclic directed dependency graph. For example, if a dependency has multiple parents in the dependency graph, it is only calculated when all its parents are already visited with settled versions, allowing accurate underlying call graphs to be considered during the calculation.}

\item \textbf{Dynamic Structure of the Dependency Graph}: 
Traditional topological sorting algorithms~\cite{Topo} operate on static graphs, processing each node once without changes during traversal. However, in our context, the dependency graph evolves as dependencies are processed, sometimes requiring multiple updates for a single node. To ensure that a node’s dependents are processed before the node itself, we modify the topological sorting algorithm. Specifically, we redefine the in-degree of a node to represent the number of computed dependents rather than just direct dependents. This dynamic in-degree update ensures that even as the graph changes, a dependency with zero in-degree—indicating all required predecessors have been processed—can still be selected for updating. This adaptation allows \tool to maintain the correct {upgrade order}, ensuring accurate dependency impact analysis as the graph evolves.
\end{itemize}

Using the traversal method described above, \tool iteratively processes nodes in the dependency graph until all are updated, ultimately producing an upgraded dependency graph that minimizes the project's technical lag as much as possible.

\subsection{Upgrading the Dependency}
When traversing a node, \tool first identifies a set of potential optimal versions as upgrade targets. {Since its primary goal is to reduce technical lag, it considers the original version and all newer versions as candidates.} Defining what constitutes a newer version is critical, as the ideal upgrade depends on various factors~\cite{cox2015measuring, gonzalez2017technical, zerouali2019formal}.
In this paper, we choose the candidate version based on SemVer, selecting the version with the highest version number, as sorting versions by SemVer~\cite{semver} can accurately reflect the evolutionary changes in the library code.

For a dependency graph 
$G$ = <$N$, $D$, $A$, $I$>, 
let $d\in N\backslash Client$ be a dependency in the graph, and the current version of {$d$} is $v(d)$, then the $version(d)$ is a set of candidate versions of $d$. $version(d)$ is denoted as ${version(d)}={\{ v' \mid v' \geq v(d), v' \text{ is a stable version of } d \}}$. 
\tool retrieves $version(d)$ from
the pre-computed database.

After determining and sorting all candidate versions, \tool calculates the technical lag by measuring the number of versions between the current and the latest version.
Then it filters the candidate versions through two iterations: first by {pruning}, then by compatibility. \textbf{Compatibility analysis} ensures safe updates by preventing breaking changes, while \textbf{pruning} improves maintainability, reduces conflicts, and prevents redundant dependencies from increasing lag. Balancing both is crucial—ignoring compatibility risks leads to breakages, while ignoring {pruning} retains unnecessary dependencies, complicating future upgrades.
The following sections will provide detailed explanations of these two iterations of version filtering during the process of upgrading.

\subsubsection{Filtering Versions by {Pruning}}\label{sec:3.5.1}
The first iteration of filtering applies the {pruning} constraint to eliminate versions that introduce unnecessary dependencies. \tool leverages the pre-computed MongoDB database storing all dependencies of libraries in MCR to assess whether upgrading to a target version increases the dependency graph size. As mentioned in~\Cref{sec:3.1.1}, this database contains the information of every Maven artifact's direct dependencies by resolving POM files.

To determine redundant dependencies, \tool computes the transitive dependencies of each version by iteratively resolving its dependencies and integrating them into the current dependency graph. 
{Given a version $v$ of dependency $d$, let $T(v)$ be the set of transitive dependencies introduced by $d$ at $v$ version. 
A version $v'$ is considered to introduce redundant dependencies if $\Delta T = T(v') - T(v_o), \text{ where } \Delta T \neq \emptyset \text{ and } \exists x \in \Delta T$, 
meaning that $\Delta T$ is a non-empty set.
$v_o$ is the original version before upgrading. $T(v') - T(v_o)$ represents the new transitive dependencies introduced by upgrading from $v_o$ to $v'$. Since these dependencies were not present in the original dependency graph, they are considered redundant.}
\tool performs this check iteratively, resolving both the dependencies declared by the target version and the transitive dependencies introduced by those dependencies, ensuring a reasonable computational cost. 
\tool filters out versions that introduce {redundant} dependencies, ensuring that the dependency graph does not grow unnecessarily.

\subsubsection{Filtering Versions by Compatibility}\label{sec:3.5.2}
The second iteration of filtering applies the compatibility assurance. 
To accurately determine which API changes may be breaking, \tool analyzes not only method invocations but also all possible usages of APIs, leveraging points-to analysis. Unlike prior work such as UPCY~\cite{dann2023upcy} and GoblinUpdater~\cite{jaime2024balancing}, which primarily focus on direct invocations, \tool considers a broader range of API interactions, ensuring a more comprehensive compatibility assessment.
To this end, \tool constructs two types of graphs as follows:

\begin{itemize}[leftmargin=5pt]
\item \textbf{Call Graph (CG)}: To analyze method reachability, \tool employs Soot’s SPARK algorithm~\cite{spark} to derive the CG from class files within the JAR package. This captures direct and indirect method calls, providing insights into how methods are invoked within the dependency graph.
\item \textbf{Class Dependency Graph (CDG)}: While the CG captures method calls, it does not account for cases where a class depends on another class without explicitly invoking its methods. To address this problem, \tool introduces a new graph named class dependency graph, where nodes represent classes in the dependency, and edges indicate type-level dependencies. These dependencies arise when a class references another class in its methods or fields, extends a superclass, or implements an interface. 
\end{itemize}
Unlike a traditional class hierarchy graph, the CDG captures concrete type interactions, allowing \tool to identify API usages beyond explicit method invocations. Since client-impacting breaking changes can arise not only from method changes but also from modifications to class hierarchies or field accesses, the CDG is essential for identifying potential breakages that using a call graph alone can not capture.

Instead of parsing the Uber JAR~\cite{uberjar} (which would be computationally expensive because the Uber JAR typically has a large size~\cite{zhang2023compatible}), \tool incrementally constructs these graphs. It first generates separate {call graphs (CGs)} and {class dependency graphs (CDGs)} for each dependent JAR package and then incrementally builds the complete CG and CDG starting from the client project.  {This modular approach aims to improve efficiency while maintaining accuracy.}

To assess API reachability within a dependency, \tool first extracts all APIs from the dependency’s bytecode using the Byte Code Engineering Library (BCEL)~\cite{bcel}. It then matches these extracted APIs with the CG and CDG of the dependency’s predecessor nodes (i.e., its dependents) to determine API usage. Specifically, \tool identifies caller-callee relationships by matching the callees in the CG and CDG against the APIs extracted from BCEL, thereby determining which APIs in the dependency are actually being called. The CG captures direct method invocation chains, while the CDG expands this analysis by incorporating additional usage relationships, such as field accesses and class inheritance.

Once the callees in the dependency are identified, \tool{} further traverses the CG and CDG of the dependency itself, using these callees as entry points. This step ensures that all reachable APIs within the dependency are identified, not just those directly matched to the dependents. Only if these reachable APIs introduce client-impacting breaking changes during an upgrade do they impact the overall project compatibility.

After determining the reachable API of a dependency, \tool proceeds to identify potential incompatible API changes that may occur during version transitions. Specifically, \tool relies on a static compatibility checker and its own reachability analysis to determine the compatibility regardless of the SemVer. For this purpose, \tool uses Revapi, a tool capable of comparing two versions of a JAR package and detecting API changes that may break compatibility. Since \tool computes the reachable API for different dependents, it also assesses the compatibility of a version with each dependent individually. When judging whether a version is compatible with a specific dependent, \tool compares this version with the dependent's version using Revapi to obtain a list of breaking API changes between the two versions. If any of the reachable API from this dependent are present in the breaking API list, \tool flags the version as incompatible. A version is deemed compatible only if it is compatible with all its dependents. During the second iteration of filtering, \tool eliminates all versions that introduce incompatibility issues.

\subsubsection{Version Selection}
The remaining versions are considered compatible and do not introduce any unnecessary dependencies. Therefore, \tool selects the latest version among them and flags it as the best version, which becomes the target version when updating the dependency. After upgrading the dependency, the structure of the dependency graph may change. As a result, \tool updates the graph accordingly, which will be described in detail in the next section.
{According to a recent study~\cite{zhang2023mitigating}, 99\% of the version declarations are not ranged in Maven. To handle the version ranges,  \tool selects the newest versions among the ranges
following what Maven does~\cite{MavenVersion}.}

\subsection{Updating the Dependency Graph}

{Upgrading} a dependency can alter the structure of the dependency graph, as the node representing the dependency might have different successors, or might depend on different versions of those successors after the update.

{To ensure the accuracy of subsequent calculations, \tool updates the graph based on the direct successor nodes both before and after the upgrade. To distinguish between the two behaviors, in this paper, we refer to the behavior of refreshing the dependency graph as ``update'', and the act of changing the version of a single dependency as ``upgrade''.}
The pre-upgrade successor nodes can be retrieved from the current dependency graph, and \tool uses a pre-built MongoDB database to query the new successor nodes, including their groupId, artifactId, and version. To ensure that only compile and runtime dependencies remain in the graph, \tool removes any dependencies whose scope is not set to runtime or compile after querying the new dependencies. Furthermore, optional dependencies are removed unless they are direct dependencies of the client project.

In conclusion, \tool was designed to reduce the technical lag of the client project. It first restores the dependency graph, then traverses the graph and upgrades every node as much as possible. For each node, {\tool filters out the versions that} introduce {redundant} dependencies or break the project, then upgrades the node to the newest version among the remaining versions. After upgrading a node, \tool updates the dependency graph accordingly to reflect its up-to-date status.

\section{Evaluation}
We aim to answer the following research questions:

\noindent{\textbf{RQ1:} How effective is \tool compared with baselines?}

\noindent\textbf{RQ2:} What impact does each component of \tool have on the overall performance?

\noindent\textbf{RQ3:} 
How does the upgrade of transitive dependencies affect the compatibility of the client?

\subsection{Preparation of Dataset}
{We selected representative repositories and modules from GitHub: 
First, we focused on Java repositories that use Maven for dependency management. We ranked these repositories by the number of stars to prioritize widely used projects. We selected popular repositories because they are widely used, and thus, technical lag has a greater impact on users. Moreover, these repositories typically maintain dependencies well, so evaluating them can reflect the tool's impact and effectiveness.
We rolled back these repositories to their previous tags to ensure they are in a stable state. 
Next, we filtered out repositories that could not be built, failed tests, or whose dependency tree could not be generated using Maven commands. 
Because adapting running environments for various individual projects to compile and test the upgraded modules for evaluation is time-consuming,
after this filtering step, we retained \textbf{15 repositories with 356 modules} that compiled successfully, passed tests, and had analyzable dependency graphs. These repositories were chosen based on stable tags or releases, ensuring a realistic dependency structure. }

A module is the basic building unit of Maven projects, so a module is a client~\cite{module}. At the module level, our dataset has a similar size to previous works. For example, Jaime et al.~\cite{jaime2024balancing} used 107 single-module projects, yielding 107 modules, i.e., 107 clients, comparable to ours. Across the modules, we analyzed 9,534 dependencies, resulting in 26.68 dependencies per module, capturing real-world dependency upgrade scenarios. On average, these repositories have 48.43K stars, further reinforcing their popularity and practical relevance. These statistics demonstrate that our dataset is both large-scale and representative. So the dataset is well-suited for evaluating \tool’s effectiveness.

\subsection{Experiment Setup}
\tool is implemented in Python 3.10.12. 
The Java environment was set to Java 17, which is widely adopted and is the minimum supported version for Spring Boot 3.0, a prominent and commonly used framework for Java-based web development, ensuring compatibility with most of the high-star repositories. The Maven version was set to \texttt{v3.9.5}.
{All the experiments were carried out on a server running Ubuntu 22.04.5 LTS with 188 GB of memory and 80 logical cores (Intel(R) Xeon(R) 6248 CPU @ 2.50GHz).}

\renewcommand{\arraystretch}{1.2}
\begin{table*}
\setlength{\tabcolsep}{2pt}
\captionsetup{justification=centering, width=\textwidth}
\caption{{{Comparison of \tool among other tools.}}}\label{tab:result}
\resizebox{1\textwidth}{!}{
\begin{tabular}{c|ccccccc|ccc|cc|c}
\hline
                              & \multicolumn{7}{c|}{\#Reduced Version Lag (Versions)}                                                                                                                                                    & \multicolumn{3}{c|}{}                                                               & \multicolumn{2}{c|}{}                                   &                                  \\ \cline{2-8}
                              &                         &                         &                         & \multicolumn{1}{c|}{}                                        & \multicolumn{3}{c|}{Total}                                  & \multicolumn{3}{c|}{\multirow{-2}{*}{\#Reduced Time Lag (y/m/d)}}                   & \multicolumn{2}{c|}{\multirow{-2}{*}{\#Broken Modules}} &                                  \\ \cline{6-13}
\multirow{-3}{*}{Tool}        & \multirow{-2}{*}{Major} & \multirow{-2}{*}{Minor} & \multirow{-2}{*}{Patch} & \multicolumn{1}{c|}{\multirow{-2}{*}{Pre-release}}           & Dir.           & Tran.           & All (Avg. over mod)      & Dir.                 & Tran.                  & All (Avg. over mod)                 & Compile                    & Test                       & \multirow{-3}{*}{\#Redunt. Deps} \\ \hline
\rowcolor[HTML]{EFEFEF} 
\textbf{\tool} & \textbf{901}            & \textbf{5,577}          & \textbf{25,234}         & \multicolumn{1}{c|}{\cellcolor[HTML]{EFEFEF}\textbf{24,539}} & \textbf{4,561} & \textbf{51,690} & \textbf{56,251 (158.01)} & \textbf{292y 8m 24d} & \textbf{5,178y 5m 24d} & \textbf{5,471y 4m 13d (15y 4m 13d)} & \textbf{0}                 & \textbf{12}                & \textbf{-353}                    \\
Dependabot                    & 11                      & 512                     & 719                     & \multicolumn{1}{c|}{6,027}                                   & 517            & 6,752           & 7,269 (29.91)            & 221y 8m 11d          & 800y 2m 22d            & 1,011y 11m 3d (4y 1m 29d)           & 15                         & 25                         & 84                               \\
\rowcolor[HTML]{EFEFEF} 
Snyk                          & 2                       & 95                      & 1,866                   & \multicolumn{1}{c|}{\cellcolor[HTML]{EFEFEF}5,695}           & 1,250          & 6,408           & 7,658 (39.74)            & 167y 6m 8d           & 58y 1m 20d             & 225y 7m 28d (1y 2m 1d)              & 17                         & 24                         & -139                             \\
GoblinUpdater                 & 0                       & 0                       & 0                       & \multicolumn{1}{c|}{0}                                       & 0              & 0               & 0 (0)                    & 0                    & 0                      & 0                                   & 0                          & 0                          & 0                                \\ \hline
\end{tabular}
}
\end{table*}

\subsection{Baselines Selection}
To evaluate \tool, we selected three representative baselines that collectively reflect the state-of-the-art in reducing technical lag: one academic tool (\textbf{GoblinUpdater}~\cite{jaime2024balancing}) and two widely adopted industrial tools \textbf{Dependabot}~\cite{dependabot} and \textbf{Snyk}~\cite{snyk} (Both have indicated the ability to reduce technical lag in their documentation~\cite{dependabotRole,snykRole}).
We exclude other tools for the following reasons: {UPCY}~\cite{dann2023upcy} upgrades one dependency at a time with compatibility checks but lacks a comprehensive strategy for global lag reduction.
{CORAL}~\cite{zhang2023compatible} and {Steady}~\cite{ponta2018beyond} focus on 
vulnerability remediation, which considers both upgrades and downgrades.
{Renovate}~\cite{Renovate} provides similar functionalities to Dependabot. 
Given Dependabot's broader adoption~\cite{he2023automating}, we selected it as the representative.

\subsection{Evaluation Metrics}
We computed the following metrics to evaluate the effectiveness of tools in RQ1 and RQ2:
\begin{itemize}[leftmargin=5pt]
    \item \textbf{Reduced Time Lag and Reduced Version Lag}: 
    Our primary goal is to reduce the technical lag, which consists of {time lag (\Cref{def:TL}) and version lag (\Cref{def:VL})}. 
    To comprehensively evaluate the effectiveness of \tool, we calculated the reduction in both version lag and time lag in RQ1 and RQ2. We presented the reduced version lag at Major, Minor, Patch, and Pre-release levels, along with the dissection of Direct vs. Transitive dependencies. Additionally, the reduced time lag is quantified as a duration represented by years, months, and days (y/m/d).

\item \textbf{Compilation Failure}: The syntactic breaking changes introduced during the upgrade could affect the recompile process of the client project. In our case, a module is a client project. Therefore, the fewer modules that fail to recompile after the upgrade, the more accurate the dependency management's judgment of syntactic compatibility during upgrading.

\item \textbf{Test Failure}: Although \tool itself does not account for semantic compatibility, we introduced this metric during the evaluation process to provide a comprehensive assessment of how the upgrade impacts overall compatibility. Specifically, the number of modules that fail tests after an upgrade indicates the extent to which the upgrade introduces issues affecting the functionality of client projects~\cite{jayasuriya2024understanding}. In our case, each module represents a client project. Therefore, the more modules that fail tests after the upgrade, the greater the likelihood that the upgrade has caused functional regressions or incompatibilities.

\item \textbf{{Redundant Dependencies (Redunt. Deps)}}: During the upgrade process, if redundant dependencies are introduced, the total number of dependencies in the dependency graph will increase. Therefore, we calculated the number of added dependencies in the graph after the upgrade to reflect the effect of {pruning}. The fewer the added dependencies, the better the {pruning} effect.
\end{itemize}

\subsection{RQ1: Effectiveness Analysis}\label{sec:4.5}
Evaluation of effectiveness has been conducted on four tools regarding the technical lag reduction, compatibility, and the redundant dependencies in~\Cref{tab:result}.
\subsubsection{\tool}
{
\tool outperforms baselines in reducing time lag and version lag, 
reducing total version lag by 56,251 versions and time lag by 5,417 years, 4 months, and 15 days,
far surpassing Dependabot and Snyk. 
In addition, the reduced version lag at all levels is greater than other tools at all levels. 
Notably, \tool reduced 901 version lag across major upgrades, suggesting that it is not much constrained by semantic versioning rules and thus more effective in reducing technical lag. 
This is attributed to \tool's ability to upgrade both direct and transitive dependencies, ensuring the latest versions are selected without breaking compatibility. 
\tool reduced 11 times more lag for transitive dependencies than the direct, which significantly outperformed Dependabot and Snyk, because \tool supports addressing transitive dependencies to effectively lower overall technical lag.
{It took around 36 hours to complete all modules, an average of 367 seconds per module, indicating an efficient online phase for individual modules.}
}

{
As for compatibility and redundancy pruning, since \tool focuses on syntactic compatibility with only mitigation for the semantic compatibility, 12 test failures emerged after the upgrade, caused by API behavioral changes not captured by syntactic analysis.
Furthermore, \tool successfully reduced 353 dependencies across the dataset by selecting versions that introduce fewer dependencies. 
\tool not only prevents redundant dependencies but also removes existing ones from the dependency graph, contributing to the overall reduction.
It is reasonable as pruning, leading to the removal of redundant dependencies, reduces graph size, and helps reduce technical lag (including time lag and version lag). Thus, the removed 353 redundant dependencies can contribute to the total reduction. We have separately calculated the contribution of pruning and found that the removed redundant dependencies only accounted for 21.2\% of the total time lag (1,171 years, 6 months, and 21 days) and 23.15\% of the total version lag (13,023). This indicates that \tool still made a significant effort to reduce technical lag by upgrading to non-breaking versions.
}

{
In the case of \Cref{sec:2.4}, \tool effectively upgrades the outdated dependencies in the \textit{spring-cloud-starter-stream-rocketmq}
module. After upgrading with \tool, 35 have been completely upgraded to the latest, with the other 85 partially upgraded, while still being able to recompile and pass all tests. Additionally, three dependencies experienced a reduction in major-level version lag, indicating that, beyond strictly following SemVer, \tool can further reduce technical lag.
}

\subsubsection{Dependabot}
{Compared to \tool, Dependabot achieved significantly less reduction in both version lag and time lag. This is because Dependabot primarily focuses on direct dependencies, and indirect dependencies are only upgraded passively alongside the direct dependencies. Because transitive dependencies are not considered, the compatibility between direct and transitive dependencies remains risky for breaking changes due to the passive upgrades. In case of version conflicts of the transitive dependencies, incompatibility would emerge.
Accordingly, \textbf{Dependabot caused 15 module compilation failures and 25 module test failures after upgrades}. Another reason for the incompatibility is its reliance on compatibility scores derived from other projects' issues rather than assessing the actual impact on compatibility~\cite{dependabotRole}. 
}

\subsubsection{Snyk}
{As Snyk has another goal of fixing vulnerabilities, it has very conservative suggestions for non-vulnerable dependencies. Therefore, it is expected that Snyk achieves less technical lag reduction compared to \tool. Notably, during the upgrade process of Snyk, direct dependencies reduced more time lag (167 years) than transitive ones (58 years), but less version lag, caused by the large time gap between version releases for certain dependencies.
Thus, analyzing technical lag from both version and time perspectives offers a more comprehensive view. Additionally, since this tool does not consider compatibility based on code during upgrades, Snyk results in 17 compilation failures as well as 24 testing failures. Notably, Snyk can reduce the number of dependencies in the dataset by 139, suggesting support for {pruning}. 
}

\subsubsection{GoblinUpdater}
{
To avoid bias, we set the maximum time limit for solving a module with GoblinUpdater~\cite{jaime2024balancing} to 80 minutes based on their recommendation. 
The tool took over 19 days, 18 hours, exceeding the time limit for all modules in our dataset, leading to no upgrade plan for any of the modules. Another reason for this failure could be attributed to the inability to handle multi-module projects, which are common in popular Maven projects. This inability has been confirmed by the authors via e-mail.
This also highlights that \tool has higher general applicability compared to GoblinUpdater, emphasizing \tool's applicability.
}

\renewcommand{\arraystretch}{1.2}
\begin{table*}
\captionsetup{justification=centering, width=\textwidth}
\caption{{Effectiveness of compatibility analysis and pruning components.}}\label{tab:Ablation}
\resizebox{0.9\textwidth}{!}{
\begin{tabular}{c|ccccc|c|cc|c}
\hline
\rowcolor[HTML]{FFFFFF} 
\cellcolor[HTML]{FFFFFF}                       & \multicolumn{5}{c|}{\cellcolor[HTML]{FFFFFF}\#Reduced Version Lag (Versions)}                            & \cellcolor[HTML]{FFFFFF}                                             & \multicolumn{2}{c|}{\cellcolor[HTML]{FFFFFF}\#Broken Modules} & \cellcolor[HTML]{FFFFFF}                                  \\ \cline{2-6} \cline{8-9}
\rowcolor[HTML]{FFFFFF} 
\multirow{-2}{*}{\cellcolor[HTML]{FFFFFF}Tool} & Major & Minor  & Patch   & \multicolumn{1}{c|}{\cellcolor[HTML]{FFFFFF}Pre-release} & Total              & \multirow{-2}{*}{\cellcolor[HTML]{FFFFFF}\#Reduced Time Lag (y/m/d)} & Compile                          & Test                       & \multirow{-2}{*}{\cellcolor[HTML]{FFFFFF}\#Redunt.  Deps} \\ \hline
\rowcolor[HTML]{EFEFEF} 
\tool                           & 901   & 5,577  & 25,234  & \multicolumn{1}{c|}{\cellcolor[HTML]{EFEFEF}24,539}      & 56,251 (159.01)    & 5,471y 6m 18d (15y 4m 13d)                                           & {0}                       & 12                         & -353                                                      \\
\rowcolor[HTML]{FFFFFF} 
Pruning Only                                   & 909   & 5,666  & 24,186  & \multicolumn{1}{c|}{\cellcolor[HTML]{FFFFFF}27,499}      & 58,260 (163.61)    & 5,403y 8m 3d (15y 4m 29d)                                            & 0                                & 12                         & -428                                                      \\
\rowcolor[HTML]{EFEFEF} 
Compatibility Only                             & 7,976 & 47,600 & 585,202 & \multicolumn{1}{c|}{\cellcolor[HTML]{EFEFEF}36,256}      & 677,034 (1,901.78) & 34,787y 2m 7d (97y 8m 18d)                                           & 3                                & 15                         & 62,394                                                    \\
\rowcolor[HTML]{FFFFFF} 
Naive                                          & 9,076 & 500,53 & 613,471 & \multicolumn{1}{c|}{\cellcolor[HTML]{FFFFFF}38,961}      & 711,561 (1,998.77) & 35,676y 9m 10d (100y 2m 29d)                                         & 3                                & 15                         & 66,234                                                    \\ \hline
\end{tabular}
}
\end{table*}

{In summary, \tool effectively balances compatibility and {pruning} while significantly reducing technical lag, and the comparison with baselines demonstrates the effectiveness of the mechanism which explicitly operates on transitive dependencies. \tool achieved a reasonable time cost, 36 hours to handle all modules.
Dependabot and Snyk are both GitHub-integrated plug-ins that create pull requests to upgrade dependencies. Measuring their execution time is not meaningful, as the majority of the time is spent
on network latency and GitHub API communication. And GoblinUpdater took around 7 days. 
}

{We have applied the data from different tools to each project on our GitHub repository. Based on the reduced total version lag of \tool for each project, we calculated the following:
\ding{172} \textbf{Confidence Interval~\cite{wonnacott1990introductory}:} The mean reduction in technical lag was 3,750.07, with a 95\% confidence interval of [837.92, 6,662.21] versions, 
indicating 95\% confidence that the true reduction lies within this range. This suggests that \tool is likely to be effective in reducing technical lag even on other samples.
\ding{173} \textbf{Hypothesis Testing:} A Wilcoxon signed-rank test~\cite{wilcoxon1945individual} was conducted with a test statistic of 105.0 and a p-value of 0.00049, indicating statistical significance (p $<$ 0.05), 
suggesting that the improvement by \tool is not due to the sample.
\ding{174} \textbf{Effect Size:} Cliff's delta~\cite{meissel2024using} was 0.9333, suggesting a large effect size and indicating that the reduction in version lag is both statistically significant and practically important.}

{\subsubsection{False Negatives and False Positives Analysis of \tool} \label{sec:fp-fn}
We further analyze the false negatives (FNs) and false positives (FPs) of \tool against three main objectives: incompatibility, redundant dependency, and technical lag reduction.}

{
\textbf{Incompatibility Determination:} 
\ding{172} FPs occur when benign API changes are wrongly seen as incompatible, due to static analysis assuming unused APIs are reachable. \tool uses a conservative strategy to minimize breaking changes, so FPs's impact is minimal.
\ding{173} FNs are missed issues, as static analysis fails to detect some API usages. While no recompilation failures occurred after dependency upgrades, \tool caused test failures in 12 modules, linked to static analysis's inability to detect runtime behaviors, like Java reflection~\cite{bruce2020jshrink, landman2017challenges, bodden2011taming, li2016reflection}. These FNs are inevitable due to the limitations of static analysis. We built upon prior similar work~\cite{snyk,steady,dann2023upcy,jaime2024balancing,zhang2023compatible,EndorLab}, with \tool focusing on detecting syntactic {incompatibility issues} to mitigate FNs.
}

{
\textbf{Redundant Dependency Detection: }Redundant dependency detection in \tool relies on a local database of MCR data. However, the database may be outdated, causing a gap that leads to FPs and FNs in detecting redundant dependencies. This gap is hard to quantify, as continuous comparison with MCR is infeasible. Nevertheless, RQ1 results show \tool reduces the dependency graph size, with a Redundant Deps metric of -353.
}

{
\textbf{Reduced Technical Lag:} Similar to redundant dependency determination, technical lag reduction is evaluated based on version information in our MongoDB database. The inevitable temporal gap between the database and the latest MCR state can introduce FPs and FNs.
In addition, the dependency resolution used in \texttt{dependency:tree} can be configured~\cite{MvnResolveConfigured} diversely, leading to varying dependency trees and inaccuracies in the parsed tree and the retrieved dependency versions. 
Despite our efforts to analyze as many projects as possible, ensuring that the parsed tree accurately retrieves the dependency versions, some inaccuracies remain.
}

\mybox{
\textbf{Answer to RQ1:}  {
\tool achieved the greatest reduction in technical lag, outperforming Dependabot, Snyk, and GoblinUpdater without recompilation failure, and achieved pruning.
}
}

\subsection{RQ2: Ablation Analysis}\label{sec:4.6}
To ensure compatibility and avoid introducing redundant dependencies during the upgrade process, \tool performs two rounds of filtering based on the requirements for compatibility and {pruning}. To assess the effectiveness of the compatibility analysis and {pruning} process, we introduce three self-developed baselines: 
\begin{itemize}[leftmargin=5pt]
    \item \textbf{{Pruning} Only}: Selecting the newest version that does not introduce {redundant} dependencies. This baseline is a variant of \tool, where the ``Filtering Versions by Compatibility'' constraint is disabled during dependency upgrades.
    \item \textbf{Compatibility Only}: Selecting the newest version that is compatible. This baseline is a variant of \tool, where the ``Filtering Versions by {Pruning}'' constraint is disabled during the process of dependency upgrades.
    \item \textbf{Naive}: Selecting the newest version, ignoring compatibility and {pruning} requirement. This baseline is a variant of \tool, disabling the process of filtering versions and just selecting the newest versions when upgrading dependencies.
\end{itemize}
{The experiment was conducted on the same dataset used in RQ1. The results are shown in~\Cref{tab:Ablation}.}
The analysis of each component is supplied as follows: 
\textbf{{Pruning} Only}: As observed in the~\Cref{tab:Ablation}, the {Pruning} Only approach reduces fewer version lag and time lag than \tool, even though it pruned 428 more dependencies than \tool. Since our main goal is to reduce technical lag, Pruning Only cannot achieve as much reduction as \tool, indicating the inappropriateness.
    It was observed that considering {pruning} also helps maintain compatibility, 
    This can be understood as follows: if the upgraded version does not introduce additional dependencies, the changes are likely to be fewer, thus reducing the likelihood of {incompatibility issues}. 
\textbf{Compatibility Only}: The Compatibility Only approach had more reduction in technical lag than {Pruning} Only.
    However, it also introduced 62,394 new dependencies serving as redundant dependencies.
\textbf{Naive}: The Naive approach reduced the most technical lag, as this baseline updated all dependencies to the newest versions. 
    However, Naive also introduced 66,234 new dependencies, which is the most among these three baselines.

Notably, both the Compatibility Only approach and the Naive approach resulted in 3 modules failing to recompile. 
These failures are related to Maven configuration, not incompatibility issues. Specifically, the error occurs because Maven is trying to resolve the javax.xml.bind:jaxb-api dependency, but it is unable to do so due to a repository URL
\footnote{The repository http://maven.java.net uses the HTTP protocol, which Maven's default security settings block (via maven-default-http-blocker) to prevent the use of unsecured HTTP connections.} 
being blocked. 
Since our dataset consists of modules with high star counts, the dependencies are relatively recent, and as a result, {incompatibility issues} were minimal when upgrading directly to the latest versions. To demonstrate the effectiveness of the compatibility analysis component, we will use a different dataset in RQ3 to explore potential client-impacting APIs through compatibility analysis.

\mybox{
\textbf{Answer to RQ2}: {
Pruning Only effectively reduced unnecessary dependencies, strongly contrasting with Naive's addition of massive redundancy. Pruning Only also mitigates upgrade risks by maintaining a leaner dependency set. Finally, Compatibility Only's few compile failures were attributed to Maven configuration, rather than API incompatibility.
}
}

\subsection{RQ3: Distribution Analysis}
{
\tool{} advances prior work by explicitly analyzing how upgrading indirect dependencies affects the compatibility of client projects, using API reachability to quantify this impact. Although Jayasuriya et al.~\cite{jayasuriya2023understanding} acknowledged that upgrades to indirect dependencies could affect client compatibility, they did not provide a concrete method to measure this risk. Our analysis fills this gap by systematically evaluating the compatibility impact across transitive dependency upgrades. 
In practice, dependency management typically has three upgrade strategies~\cite{jaime2024balancing} based on SemVer~\cite{jaime2024balancing}:
\textbf{MMP}: upgrade to the latest available version;
\textbf{mMP}: upgrade to the latest version within the same major version;
\textbf{mmP}: upgrade to the latest version within the same major and minor version. 
We leverage \tool{}'s compatibility analysis to investigate how these strategies influence the number of client-impacting APIs introduced through indirect dependencies, and how this number varies with the depth of dependencies.
}

\subsubsection{Preparation of Dataset for RQ3}\label{sec:4.1}
{RQ1-2 evaluate the effectiveness of \tool, so a smaller dataset in RQ1-2 helps validate correctness, as the ground truth is easier to extract. However, RQ3 focuses on breaking changes in transitive dependency upgrades after confirming effectiveness. To better reflect the ecosystem, RQ3 uses a larger dataset different from the one in RQ1-2 to ensure the generality of the conclusion.} We selected Maven-based projects from Java repositories in GitHub with more than 100 stars, then randomly chose 500 repositories from this set. There are 8,061 modules in these repositories. Subsequently, we packaged them and extracted dependency trees from them by the Maven command, and filtered modules that could not succeed in these two processes, resulting in \textbf{1,529 modules}. 
This filtering was necessary because the compatibility detection component used in the experiment requires the compiled class files of the client project, which may not be available for download from the MCR.

\subsubsection{Experiment and Result}
Finally, we applied \tool's compatibility assessment component to the 1,529 modules; The result is shown in~\Cref{tab:Trans}. 
In the dataset, the max depth of transitive dependencies is 9. Based on the client-impacting API distribution across the transitive dependency depths, we could have the following findings about the impact of upgrading transitive dependencies: 

\begin{table}
\caption{{Distribution of broken clients and APIs when upgrading \textit{transitive} dependencies under mmp, mMP and MMP strategies (Excluding fully zero-valued rows).}}
\label{tab:Trans}
 \resizebox{0.8\columnwidth}{!}{
\begin{tabular}{c|ccc|ccc}
\hline
\rowcolor[HTML]{FFFFFF} 
\cellcolor[HTML]{FFFFFF}                                 & \multicolumn{3}{c|}{\cellcolor[HTML]{FFFFFF}\#Broken Clients} & \multicolumn{3}{c}{\cellcolor[HTML]{FFFFFF}\#Client-impacting APIs} \\ \cline{2-7} 
\rowcolor[HTML]{FFFFFF} 
\multirow{-2}{*}{\cellcolor[HTML]{FFFFFF}\textbf{Depth}} & \textbf{mmP}        & \textbf{mMP}       & \textbf{MMP}       & \textbf{mmP}          & \textbf{mMP}         & \textbf{MMP}         \\ \hline
\rowcolor[HTML]{EFEFEF} 
2                                                        & 52                  & 184                & 264                & 98                    & 562                  & 2,128                \\
\rowcolor[HTML]{FFFFFF} 
3                                                        & 17                  & 108                & 148                & 81                    & 273                  & 507                  \\
\rowcolor[HTML]{EFEFEF} 
4                                                        & 3                   & 32                 & 46                 & 52                    & 135                  & 417                  \\
\rowcolor[HTML]{FFFFFF} 
5                                                        & 0                   & 32                 & 46                 & 0                     & 135                  & 417                  \\
\rowcolor[HTML]{EFEFEF} 
6                                                        & 0                   & 6                  & 10                 & 0                     & 36                   & 245                  \\ \hline
\end{tabular}
}
\end{table}

\begin{itemize}[leftmargin=5pt]
    \item Updating transitive dependencies potentially breaks client projects, even following SemVer.
According to SemVer, both the mmP and mMP strategies should not introduce any breaking changes, due to the consistent major version. However, in our dataset, there are still 1,372 client-impacting APIs introduced through mmP or mMP to upgrade transitive dependencies.
So the techniques like \tool, which can detect the client-impacting API in even transitive dependencies, are necessary.
    \item When updating transitive dependencies, the MMP strategy, which updates to the latest versions, may introduce the most client-impacting API. 
According to the data in our dataset, there are 231 client-impacting APIs introduced through the mmP strategy, 1,141 client-impacting APIs introduced through the mMP strategy, and 3,714 client-impacting APIs introduced through the MMP strategy.
Furthermore, the more versions are updated, the greater the potential for introducing client-impacting APIs.
    \item The higher the dependency depth, the fewer the client-impacting APIs introduced. In our dataset, at depth 2, there are 2,128 client-impacting APIs introduced through MMP, while at depth 6, this number drops to 245.
    \item For transitive dependencies with depth more than 6, updating is less likely to break the client projects. No client-impacting APIs or breaking changes were introduced for dependencies beyond the 6th layer in our dataset.
\end{itemize}

Building on the findings of Jayasuriya et al.~\cite{jayasuriya2023understanding}, our study provides further quantitative insights into how clients directly use APIs from transitive dependencies. 
We quantitatively assessed the impact of transitive dependency updates on client-facing APIs, thereby filling the gap left by Jayasuriya et al.~\cite{jayasuriya2023understanding}, who did not explore this aspect in their study of clients’ direct API usage in transitive dependencies. 
We observe that only 0.06\% of clients directly invoke APIs located beyond the 6th level of transitive dependencies. This suggests that updates affecting APIs deeper than the 6th level are unlikely to introduce incompatibility issues for client software. This offers a more comprehensive understanding of the risks associated with transitive dependency updates.
\mybox{
\textbf{Answer to RQ3}: {
Upgrading transitive dependencies is more likely to break the client project when there are fewer dependency layers; dependencies beyond six layers rarely introduce client-impacting breaking changes. Furthermore, the MMP strategy (upgrading to the latest version) is more prone to client breakage than the mmP and mMP strategies.
}
}

\section{Threats to Validity}
When performing API reachability analysis, \tool uses static analysis techniques to extract static call graphs and class dependency graphs. However, static analysis is constrained by its inherent limitations, leading to FPs and FNs, particularly with dynamic behaviors like Java reflection~\cite{bruce2020jshrink, landman2017challenges, bodden2011taming, li2016reflection}, where call relationships are often undetectable. While dynamic execution (testing) can yield actual call graphs, relying solely on it is infeasible due to two main reasons: the difficulty in achieving sufficient test coverage, and the inefficiency of repeatedly executing tests to dynamically update the dependency and call graphs. To mitigate coverage limitations and enable real-time updates of the call and class dependency graphs, we employ static analysis as the necessary trade-off. 
Secondly, the datasets used in our study may limit generalizability.
To mitigate this threat, we select highly starred and widely used projects for RQs 1 and 2, and apply random yet representative sampling for RQ3 to capture diversity across project types and dependency structures. 
Thirdly, since exploring all combinations of dependency versions is computationally infeasible, our sequential upgrade strategy may not always achieve the globally minimal technical lag.
To mitigate this threat, we adopt a greedy heuristic that prioritizes stable and up-to-date versions, striking a balance between optimization quality and practical efficiency, as supported by the evaluation results.

\section{Related Work}
\subsection{Dependency Management}
Several research efforts focus on optimizing dependency management for client projects.
Dann et al.~\cite{dann2023upcy} proposed UPCY, which upgrades a dependency while minimizing breaking changes, but it is limited to single dependencies and does not assess actual impact on the client project~\cite{jayasuriya2023understanding,ochoa2022breaking}. Jaime et al.~\cite{jaime2024balancing} introduced GoblinUpdater, which updates all dependencies without breaking the client project, yet it cannot fully analyze transitive dependencies or handle multi-module Maven projects. Steady~\cite{ponta2020detection, ponta2018beyond} adjusts dependency versions to mitigate vulnerabilities but only provides a probabilistic compatibility assessment. CORAL~\cite{zhang2023compatible} upgrades vulnerable dependencies using API reachability analysis but focuses on security rather than technical lag and excludes class-level API analysis. He et al.~\cite{he2023automating} examined Dependabot, highlighting its inability to guarantee compatibility. Unlike \tool, these approaches fail to simultaneously address technical lag, compatibility, and {pruning}.
\subsection{Technical Lag and Compatibility}
Technical lag {refers to} the outdateness of dependencies and has been widely studied~\cite{hu2024empirical, zerouali2019formal, wang2020empirical, decan2018evolution, stringer2020technical, gonzalez2017technical}. Gonzalez-Barahona et al.~\cite{gonzalez2017technical} introduced the concept, while Zerouali et al.~\cite{zerouali2019formal} and Wang et al.~\cite{wang2020empirical} proposed formulas for measuring lag. Studies on NPM~\cite{decan2018evolution} and multiple package managers~\cite{stringer2020technical} highlighted its prevalence and potential mitigation. 
Compatibility research focuses on empirical studies and API compatibility checking. Empirical studies analyzed breaking changes~\cite{dig2006apis, dietrich2014broken, xavier2017historical} and their impact on clients~\cite{jayasuriya2023understanding, jayasuriya2024understanding}. Static analysis tools~\cite{japicmp, japicc, sigtest, revapi, japitools, brito2018apidiff} detect breaking changes but often yield false positives due to analyzing entire libraries rather than client-specific usage.  Maracas~\cite{ochoa2022breaking} improves precision but struggles with inheritance, overrides, and exception propagation, and overlooks indirectly invoked APIs. \tool overcomes these limitations by incorporating API call relationships, enabling a more accurate assessment of compatibility impacts.

\section{Conclusion}
In this paper, we proposed \tool to minimize the technical lag of Maven projects without introducing incompatibility or redundant dependencies. The evaluation demonstrated that \tool achieved the greatest reduction in technical lag among similar dependency management tools while ensuring successful recompilation with even smaller dependency graphs. 
A large-scale study in real-world applications found that breaking changes in APIs within transitive dependencies can also impact the client.
Besides technical lag, \tool's objectives can be flexibly extended to incorporate metrics like security to guide upgrades while avoiding {incompatibility issues} and dependency redundancy in future work.

\begin{acks}
This work was supported by the Fundamental Research Funds for the Central Universities, JLU, the Shanghai Trusted Industry Internet Software Collaborative Innovation Center, the National Research Foundation Singapore, and the Cyber Security Agency under the National Cybersecurity R\&D Program (NCRP25-P04-TAICeN).
\end{acks}

\bibliographystyle{ACM-Reference-Format}
\bibliography{reference}

\end{document}